\documentclass[preprint,showpacs,preprintnumbers,amsmath,amssymb]{revtex4}
\usepackage{graphicx}% Include figure files
\usepackage{dcolumn}% Align table columns on decimal point
\usepackage{bm}% bold math

\def\mib#1{\mbox{\boldmath $#1$}}

\begin{document}
%\preprint{arXiv:cond-mat/}

\title{
Isostaticity and Mechanical Response of Two-Dimensional Granular Piles}

\author{Akihiro Kasahara and Hiizu Nakanishi}

%\thanks{E-mail address: naka4scp@mbox.nc.kyushu-u.ac.jp}}

\affiliation{
Department of Physics, Kyushu University 33, Fukuoka 812-8581, Japan
}

%-------------------------------------------------------------
\begin{abstract}
We numerically study the static structure and the mechanical response of
two-dimensional granular piles.  The piles consist of polydisperse disks
with and without friction.  Special attention is paid for the rigid
grain limit by examining the systems with various disk elasticities.  It
is shown that the static pile structure of frictionless disks becomes
isostatic in the rigid limit, while the isostaticity of frictional pile
depends on the pile forming procedure, but in the case of the infinite
friction is effective, the structure becomes very close to isostatic in
the rigid limit.
The mechanical response of the piles are studied by infinitesimally
displacing one of the disks at the bottom.  It is shown that the total
amount of the displacement in the pile caused by the perturbation
diverges in the case of frictionless pile as it becomes isostatic, while
the response remains finite for the frictional pile.  In the
frictionless isostatic pile, the displacement response in each sample
behaves rather complicated way, but its average shows wave like propagation.
\end{abstract}


%\kword{
%static structure of granular pile, isostaticity, marginal rigidity,
%mechanical response of granule pile
%}

%-------------------------------------------------------------
%\begin{document}
\maketitle

%------------------  Introduction ----------------------------------
\section{Introduction}

A pile of granular material is often modeled as an assembly of rigid
particles.  This simple picture, however, causes some conceptual
difficulties when one consider force distribution, or stress, in the
pile.

In an ordinary solid that consists of atoms, which are deformable
elements, the stress inside is determined by the stress balance
equations with the constitutive relation between the stress and strain.
No deformation is allowed, however, in rigid elements, in which case the
forces acting on each element cannot be determined uniquely for a given
configuration in general.

There is a special class of stable pile structures called isostatic, or
marginally rigid: A pile structure is isostatic when the forces acting
between the elements are uniquely determined only from externally
applied forces and the pile structure without any information on
deformation of the elements.  For such a pile, the total number of
balance equations for the forces and torques acting each particle should
be equal to the number of independent components of forces, which means
the average coordination number $z$ is $2d$ for frictionless spheres and
$d+1$ for frictional spheres or aspherical particles in $d$-dimensions
if the pile is isostatic.

In a real pile the isostaticity is not always satisfied; in the
overconstrained case, where the number of conditions is larger than the
number of forces, the pile is unstable because there are no sets of
force that satisfy the balance equations.  On the other hand, in the
underconstrained case, where the number of conditions is smaller than the
number of forces, the forces cannot be determined uniquely from the
macroscopic structure information of the pile; the macroscopic friction
force depends on the piling history.
It has been conjectured that a stable pile of frictionless rigid
particles forms an isostatic structure\cite{M98}.

The isostaticity of a pile structure has been tested numerically and
experimentally for both the frictionless and frictional cases by
counting the coordination numbers.
Makse {\it et al.}\cite{MJS00} have performed numerical simulations for three
dimensional sphere systems without the gravity and made compact
aggregates of balls by compressing the systems by pushing the
surrounding walls.  By examining the coordination numbers in the
zero-pressure limit, they concluded both the frictionless and frictional
sphere systems become isostatic in the rigid limit.
On the other hand, Silbert {\it et al.}\cite{SEGHL02} have also performed
numerical simulations on the three-dimensional sphere system, but the
way they made piles is different from that of Makse {\it et al.}
They released the particles in the system at once under the gravity and
waited until all the particles stop.  Their conclusion is that the
frictionless pile becomes isostatic in the rigid limit, but the structure
of frictional pile depends on the piling procedure and never becomes
isostatic.
Blumenfeld and Ball\cite{BB02,BB03} have done simple table top
experiments on the pile of two-dimensional non-circular grains made of
cardboard.  Piles are formed by collecting the grains scattered
initially on a horizontal surface by sliding an open rectangular frame.
They found that the higher the starting density of grains is, the more
cooperative reconfiguration is taken place before they are stuck with
each other, and the lower the ending pile density.  They have concluded
that the isostatic structure of frictional grain is achieved in the
limiting case where the starting and ending density coincide.

The isostatic structure, if it is realized in a real pile, should be
reflected in the mechanical properties of the rigid granular pile.  The
mechanical properties of frictionless isostatic structure has been
studied in some detail, and it has been shown that 
(i) the force chain may be regarded as propagating unidirectionally\cite{TW99},
(ii) there is a correspondence between the force-force response and the
displacement-displacement response\cite{M01},
and (iii) the piles are very sensitive to an external perturbation\cite{M98};
These features may correspond with some properties of frictionless
rigid granular pile.

Regarding the mechanical response of the isostatic pile,
using a simple lattice model of isostatic structure,  Moukarzel\cite{M98} has
shown that the total response to an external perturbation diverges as
the system approaches to the isostaticity due to the ``pantograph
effect''.
As for the effects of friction, Moukarzel {\it et al.}\cite{MPRV03} have
performed experiments on the two-dimensional rigid disk system and
demonstrated that the displacement response function has the single peak
Gaussian shape with diffusive broadening.  They also performed computer
simulations on the two-dimensional frictionless disk system and showed
that the displacement response function have a double peaked shape with
the wave like propagation, suggesting the friction plays an important
role in the mechanical response.

In this paper, we present the detailed results of our numerical
simulations\cite{KN04} on the two-dimensional frictionless and
frictional disks to show how the isostatic piles are formed for
frictionless and frictional disks.  We also investigate the mechanical
response of the isostatic piles to an external perturbation, and
demonstrate there exist important differences in the response between
the frictionless and frictional piles when they approach the
isostaticity.
After introducing the model and piling procedures in Sec. 2,
we present the results for
the static
pile structure for frictionless and frictional disks in
Sec. 3.  The mechanical responses are investigated in Sec. 4.  The
summary of our results is given in Sec. 5.

%--------------- Model ------------------

\section{Model}

We perform molecular dynamics simulations on the system that consists of
two-dimensional disks with linear elasticity and damping, which is
usually called DEM(Discrete Element Method) in the engineer community.
The system is polydisperse with a uniform distribution in the disk
diameter over the range between $0.9\sigma_0$ and $\sigma_0$, with
$\sigma_0$ being the maximum diameter.  The masses of the disks are
assumed to be proportional to their areas: the mass of the disk with the
diameter $\sigma_0$ is denoted by $m_0$.  The bottom of the system is
made rough by attaching the disks with the interval $\sigma_0$, and we
employ the periodic boundary condition in the horizontal direction.  The
number of disks $N$ in the system is typically 400, and the horizontal
length of the system is $20\sigma_0$, thus we have approximately 20
layers of disks on average.

Piles are formed by letting the system run under gravity with the
acceleration $g$ from initial configurations until all the disks stop
moving.

The disk at the position $\mib{x}(t)$ with mass $m$ follows the
Newtonian equation
\begin{equation}
m\ddot{\mib{x}}(t)=\mib{F}(t),
\label{Newton}
\end{equation}
where the force $\mib{F}(t)$ consists of the gravitational and contact
forces from the neighboring disks in contact.  We also use the ``viscous
equation'' 
\begin{equation}
\gamma\dot{\mib{x}}(t) = \mib{F}(t)
\label{viscous}
\end{equation}
for some cases to examine the effect of friction because the system that
follows eq. (\ref{viscous}) is stuck as soon as the force balance is
achieved and is more affected by the friction.

The two particles
$i$ and $j$ at $\mib{x}_i(t)$ and $\mib{x}_j(t)$ with radii $r_i$ and
$r_j$, respectively, are in contact when the overlap $\delta_{ij}$ given by
\begin{equation}
\delta_{ij}=r_i+r_j-|\mib{x}_i-\mib{x}_j|
\label{delta}
\end{equation}
is positive.  Then the particle $j$ exerts the force $\mib{F}_{ij}$  on
the particle $i$
\begin{equation}
\mib{F}_{ij} = \mib{F}_{ij}^n  +\mib{F}_{ij}^t ,
\end{equation}
where $\mib{F}_{ij}^n$ and $\mib{F}_{ij}^t$ are the normal and
tangential components of the force:
\begin{eqnarray}
\mib{F}_{ij}^n & = & k_n\delta_{ij}\mib{\hat{n}}_{ij}-\gamma_n \mib{v}_{ij}^n ,
\label{normal}
\\
\mib{F}_{ij}^t & = & -k_t\Delta s_{ij}\mib{\hat{t}}_{ij}-
   \gamma_t \mib{v}_{ij}^t .
\label{tangential}
\end{eqnarray}
Here, $\mib{\hat n}_{ij}$ and $\mib{\hat t}_{ij}$ are the normal and
tangential unit vectors, respectively; $\Delta s_{ij}$ is the tangential
displacement of the contact points after the contact, $\mib{v}_{ij}^n$
($\mib{v}_{ij}^t$) is the normal (tangential) relative velocity, $k_n$
($k_t$) is the normal (tangential) elastic constant, and $\gamma_n$
($\gamma_t$) is the normal (tangential) damping constant.  Note that we
assume no threshold for the disks to slip during the contact in
eq.(\ref{tangential}), which corresponds to the case with the infinite
friction coefficient.  In the case of the frictionless disk, we simply
set $\mib{F}_{ij}^t=\mib{0}$.

%--------------------------------------------

In the actual simulations, we use 
$\gamma_n=2\sqrt{k_n} \,[\sqrt{m_0}]$ for the frictionless case, and
$k_t=0.2 k_n$ and $\gamma_n=\gamma_t=2\sqrt{k_n} \,[\sqrt{m_0}]$ for the
frictional case.  In the simulations with the viscous equation
(\ref{viscous}), we take $\gamma=5[m_0\sqrt{g/\sigma_0}]$ with
$\gamma_n=\gamma_t=0$ in eqs.(\ref{normal}) and (\ref{tangential}).

%-------------------------------------------------------------------
\section{Static Structure of Piles}

First, we study the isostaticity of granular pile formed through several
procedures.

Isostaticity is most easily checked by counting the average number of
particles in contact with each particle, namely the average coordination
number $z$.  In order that the force acting between particle can be
determined only by the contact network structure, the number of
independent components of force should be equal to the number of force
and torque balance equations.  In the case of frictionless spherical
grain, the number of independent force components is $zN/2$ because the
contact forces have only radial component.  The torque of each particle
always balances and the number of force balance equation is $Nd$,
therefore, we have $z=2d$ for the isostatic structure of
frictionless spheres in $d$-dimensions.
In the case of frictional grain, we have $z=d+1$ for both spherical and
non-spherical grain, because the number for independent force components
is $zdN/2$ and the number of the force and torque balance equations are
$dN$ and $d(d-1)N/2$, respectively.

\begin{figure}[b]
\begin{center}
\includegraphics[width=10cm,clip]{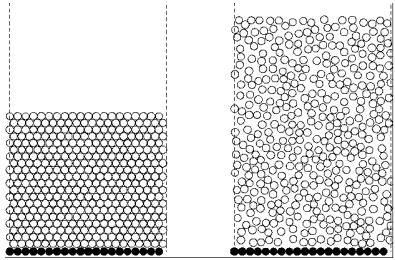}
\end{center}
\caption{
Two initial configurations:  the triangular lattice with polydisperse
 disks (left) and the random configuration with a number density 0.6
 (right).
}
\label{f-1}
\end{figure}

We perform molecular dynamics simulations to construct piles of
frictionless and frictional disks using several procedures.  We try two
types of initial configurations: 
the triangular lattice and the
random configuration (Fig. \ref{f-1}); the triangular lattice with the
lattice constant $\sigma_0$ is not a regular lattice because the disks
located at the lattice points are polydisperse.  The random
configurations are prepared by randomly arranging disks with an area
fraction of approximately 0.6.

Simulations to form piles start from these configurations with zero
particle velocity and finish when the kinetic energy of each disk
becomes negligibly small, namely, smaller than $10^{-15}[m_0g\sigma_0]$.
The number of disks $N$ is 400 and the system size is 20$\sigma_0$ in
the horizontal direction, thus the number of layers in depth is 20 on
average.  We also try both eqs. (\ref{Newton}) and (\ref{viscous}) for
the time development.

\begin{figure}[tb]
\begin{center}
\includegraphics[width=8cm]{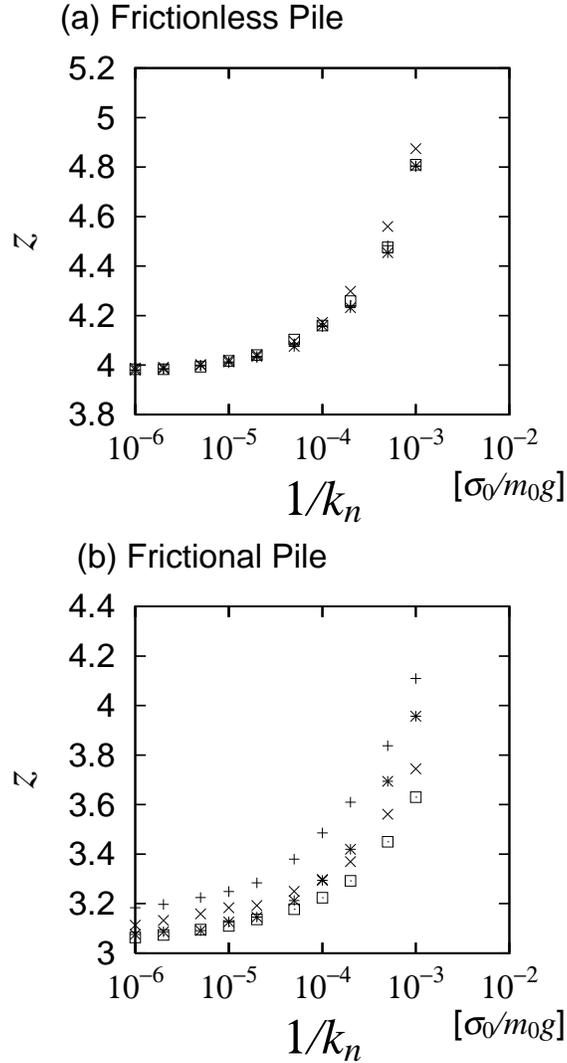}
\end{center}
\caption{ The coordination number $z$ for various elastic constants $k_n$
for a frictionless pile (a) and frictional pile (b).  The marks represent
the pile preparation procedure: the Newtonian equation with the
triangular lattice initial configuration ($+$), the Newtonian equation
with the random initial configuration ($\times$), the viscous equation
with the triangular lattice initial configuration ($*$), and the viscous
equation with the random initial configuration ($\square$).  Each mark
represents the average of six to twelve realizations.  } \label{f-2}
\end{figure}

\begin{figure}[tb]
\begin{center}
\includegraphics[width=8cm]{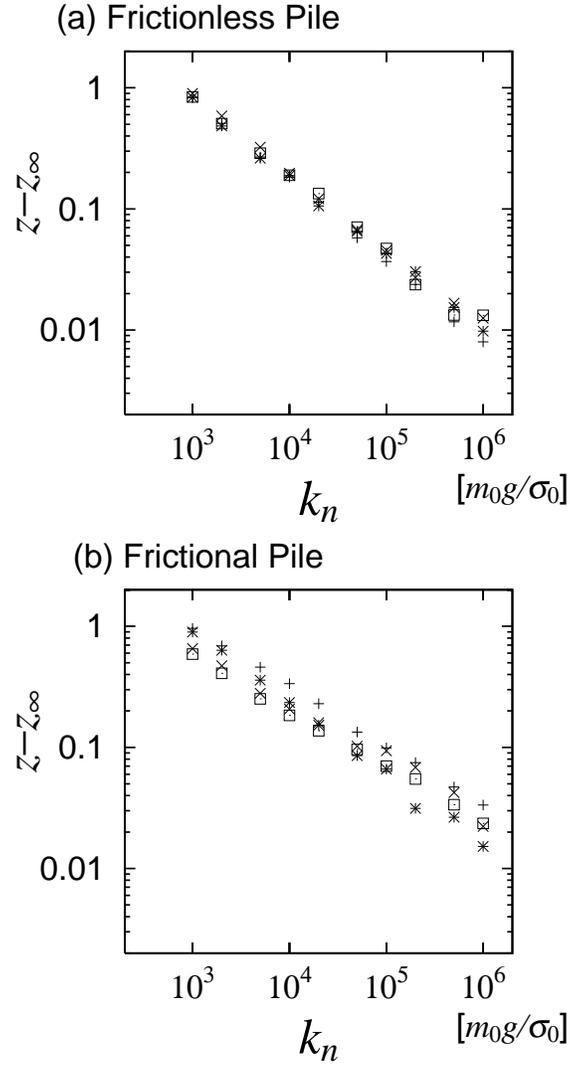}
\end{center}
\caption{ The coordination numbers $z-z_\infty$ for various elastic constants
$k_n$ in the log-log scale for the frictionless pile (a) and the
frictional pile (b).  The same data are plotted using the same marks as
those in Fig. \ref{f-2} with $z_\infty$ listed in Table \ref{t-1}.
 } \label{f-3}
\end{figure}

Figure \ref{f-2}(a) shows the $k_n$-dependence of the coordination
number $z$ for the frictionless disks.  It can be seen that the results
does not depend on the preparation procedures very much and $z$
converges to a number very close to 4 in the large $k_n$ limit for both
initial configurations and time developments.  The $k_n$-dependence is
well represented by the power law
\begin{equation}
z-z_\infty \propto k_n^{-\alpha}
\end{equation}
as is shown in Fig. \ref{f-3}(a).
The parameters $z_\infty$ and $\alpha$ are tabulated in Table \ref{t-1}.

As for the case of frictional disks, the results are shown in
Fig. \ref{f-2}(b).  There are two things to be noted in comparison with
the frictionless case: (i) the discrepancy among different preparation
procedures is larger in the frictional case, and (ii) the limiting
values of the coordination number are substantially different from 3,
{\it i.e.} the value for the isostatic structure of the frictional grain in
two dimensions(Table \ref{t-1}).

There seems to be, however, the tendency that the limiting value of the
coordination number $z_\infty$ becomes closer to 3 in the case where the
friction may produce more random pile configuration; namely from the
Newtonian with the triangular lattice initial configurations to the
viscous equation with the random initial configuration.

These results should be compared with those by Silbert {\it el
al.}\cite{SEGHL02} They also constructed granular piles using DEM and
concluded that the piles of frictionless spheres become isostatic in the
rigid limit but that of frictional spheres does not become isostatic.

The major difference between the present work and that of Silbert {\it et
al.} lies in the following points: Silbert {\it et al.}\cite{SEGHL02} studied
the three dimensional systems of mono-disperse spheres with a finite
friction constant that follows the Newtonian equation while we
investigate the two-dimensional system of polydisperse disks with the
infinite friction constant that follows the Newtonian or the viscous
equation.  Both agree in the point that the pile structure in the
frictional systems does not become isostatic in the same way as it does
in the frictionless system, but our results suggest that the
isostaticity is achieved even for a frictional system in the certain
limiting situation where the friction becomes very effective.

\begin{figure}[tb]
\begin{center}
\includegraphics[width=8cm]{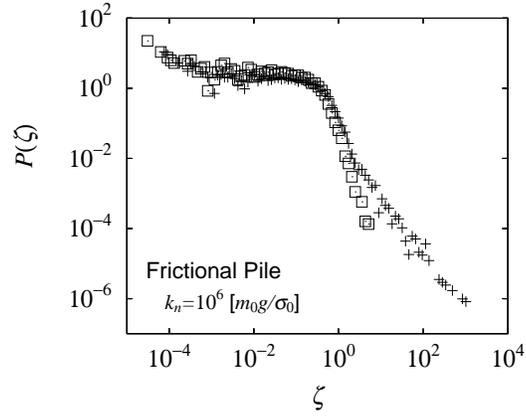}
\end{center}
\caption{ The distribution for the ratio $\zeta$ of the tangential force
 to the normal force in the piles produced by the viscous equation
 ($+$) and that by Newtonian equation ($\square$) from the random
 initial configurations with the disk elasticity
 $k_n=10^6$[$m_0g/\sigma_0$].  The average coordination numbers are
 $z=3.06$ for the pile by the viscous equation and $z=3.18$ for that by
 the Newtonian equation.  Each plot represents average over about ten
 realizations.  } \label{f-4}
\end{figure}

To examine how this discrepancy arises, we plot the distribution of the
ratio $\zeta$ of the tangential force to the normal force at each
contact.  The comparison between the pile formed via the Newtonian
equation and the pile via the viscous equation from the random initial
configurations is given in Fig. \ref{f-4} for the disk elasticity
$k_n=10^6$ [$m_0g//\sigma_0$], in which case the coordination numbers of
the piles are $z=3.18$ for the Newtonian equation and $z=3.06$ for the
viscous equation.  For the both cases we use the random initial
configurations.  One can see that the pile by the viscous equation
contains more contacts with very large value of $\zeta$, while the pile
by the Newtonian equation has only contacts with $\zeta$ smaller than 10
even though the infinite friction coefficient allows any value of $\zeta$.

If we use a finite value for the friction coefficient, some of the
contacts in the pile via the viscous equation would slip to make the
pile denser.  This would results in a larger average coordination
number, which means the pile structure becomes further away from
isostatic.

From these observations, we conclude that the infinite friction
coefficient and the viscous equation for the time development in the
pile forming process from a random initial configuration makes the
friction very effective, as a result, the pile becomes so decompacted
that it becomes nearly isostatic.

%-------------------------------------------------------------------
\section{Mechanical Response of Piles}

Now, we study the mechanical response of the frictionless and frictional
piles to the external perturbation and see how the response changes as
the pile becomes closer to the isostatic.  The perturbation is given by
displacing
 one of the disks attached at the floor by $\delta\mib{r}_0$ in the
upward direction very slowly, and we observe the displacement
of each disk $\delta\mib{r}_i$ caused in the pile by it.

In order to examine properties of a given contact network, the size of
the external displacement is taken small enough that the perturbation
does not cause any change in the connectivity of the contact network in
the pile.  In the simulation, we take the external displacement as
$\delta\mib{r_0}=(0,m_0/k_n)$, namely, the order of disk deformation.
For such small external displacement, we have checked that the contact
network in the pile does not change, and $\delta\mib{r}_i$ is
proportional to $\delta\mib{r}_0$ thus the relative displacement
$\mib{d}_i$ defined by
\begin{equation}
\mib{d}_i \equiv {\delta\mib{r}_i\over|\delta\mib{r}_0|}
\end{equation} 
does not depend on $|\delta\mib{r}_0|$.

The initial piles are prepared by the way described in the previous
section with the random initial configuration and the Newtonian
equation.
The system size is 60$\sigma_0$ in the horizontal direction and the
number of disks $N$ is 1200, thus the average number of layers in the
depth is 20.

Examples of pile response to the perturbation are shown in
Fig.\ref{f-5}.  The filled disks in the bottom layer are fixed except
one at the center marked with a red arrow, which disk is displaced
upward.  The directions and the distances of the displacements of the
disks in the pile are shown by the colored arrows: the red, green and
black arrows denotes the displacements $|\delta\mib{d}_i|\ge 1$, $0.5
\le |\delta\mib{d}_i|\le 1$, and $0.1 \le |\delta\mib{d}_i|\le 0.5$,
respectively.  The disks that move less than $0.1\delta r_0$ are not
marked.  One can see the effects of the perturbation extend over long
distance in the upper case, namely, the frictionless pile close to the
isostaticity, while the effects decay within a short distance in the
lower pile of frictional disks.

%--------------------------------------
\subsection{Total Longitudinal Response of Displacement}

As a measure of response to the perturbation, we define the total
response of displacement in the $y$-direction $D_y$, or the total
longitudinal response, as
\begin{equation}
D_y \equiv \sum_{i=1}^N \left|{\delta y_i\over\delta y_0}\right|
     = \sum_{i=1}^N |d_{iy}|,
\end{equation} 
where $\delta y_i$ is the $y$ component of the displacement for the
$i$-th disk.  This quantity should be finite if the response is confined
within a finite region, but can diverge in the infinite system if the
response extend to the infinity.

\begin{figure}[tb]
\begin{center}
\includegraphics[width=8cm]{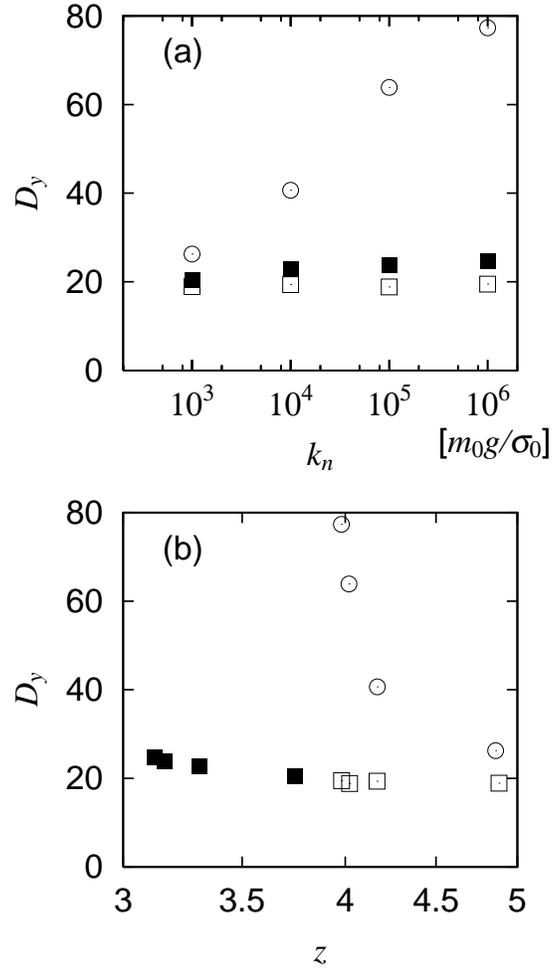}
\end{center}
\caption{ The total longitudinal displacement $D_y$ caused by the
perturbation in the frictionless pile ($\odot$), in the frictional pile
($\blacksquare$), and for the frictional response 
in the piles formed through frictionless dynamics
 ($\square$, see text).  (a) $D_y$ v.s. the elastic constant $k_n$.
(b) the same data $D_y$ are plotted v.s. the average coordination number
of the pile $z$.  Each plot represents average over around 350 realizations.
  } \label{f-6}
\end{figure}

The results are shown in Fig.\ref{f-6}(a), where the total response $D_y$ is
plotted as a function of $k_n$ with which the examined piles are formed
by the Newtonian equation from the random initial configuration.

The marks $\odot$'s ($\blacksquare$'s) in Fig.\ref{f-6}(a) denote the
total response $D_y$ of the frictionless (frictional) disks in the
frictionless (frictional) piles, respectively.

In addition to these, we examine the frictional response to the
perturbation in the frictionless piles ($\square$'s); namely, the
response is calculated using the frictional interaction between disks,
although the pile itself is prepared using the frictionless interaction,
thus the pile structure has larger coordination numbers than that of
real frictional piles with the same elastic constant $k_n$.

In Fig.\ref{f-6}(b), the same data are plotted against the coordination number
$z$ of the pile.  One can see that the total response of the
frictionless pile tends to diverge as $z$ approaches 4, while it remains
finite for the frictional pile as $z\to 3$.  It is interesting to see
that the frictional responses of the frictionless piles are almost on
the same curve with the frictional responses of the frictional piles when
they are plotted against $z$, which suggests that the coordination
number characterizes the mechanical response of the pile very well.

%--------------------------------------
\subsection{Spatial Variation of Absolute Value of Longitudinal Response}

To understand the diverging total responses in the frictionless isostatic
pile, we plot the averaged behavior of the spatial variation of
absolute value of longitudinal response $\hat d_y(\mib{r})$ defined by
\begin{equation}
\hat d_y(\mib{r}) \equiv <\sum_{i=1}^N |d_{iy}| \delta(\mib{r}-\mib{r}_i)>,
\end{equation}
where $<...>$ denotes the statistical average.  This quantity is related
to $D_y$ by
\begin{equation}
D_y = \int\int d\mib{r}  \hat d_y(\mib{r}).
\end{equation}
In actual calculations, the spatial variation is calculated on a grid
with the mesh spacing $\sigma_0$ and the average is taken over a few
hundreds realizations.

\begin{figure}[tb]
\begin{center}
\includegraphics[width=17cm,clip]{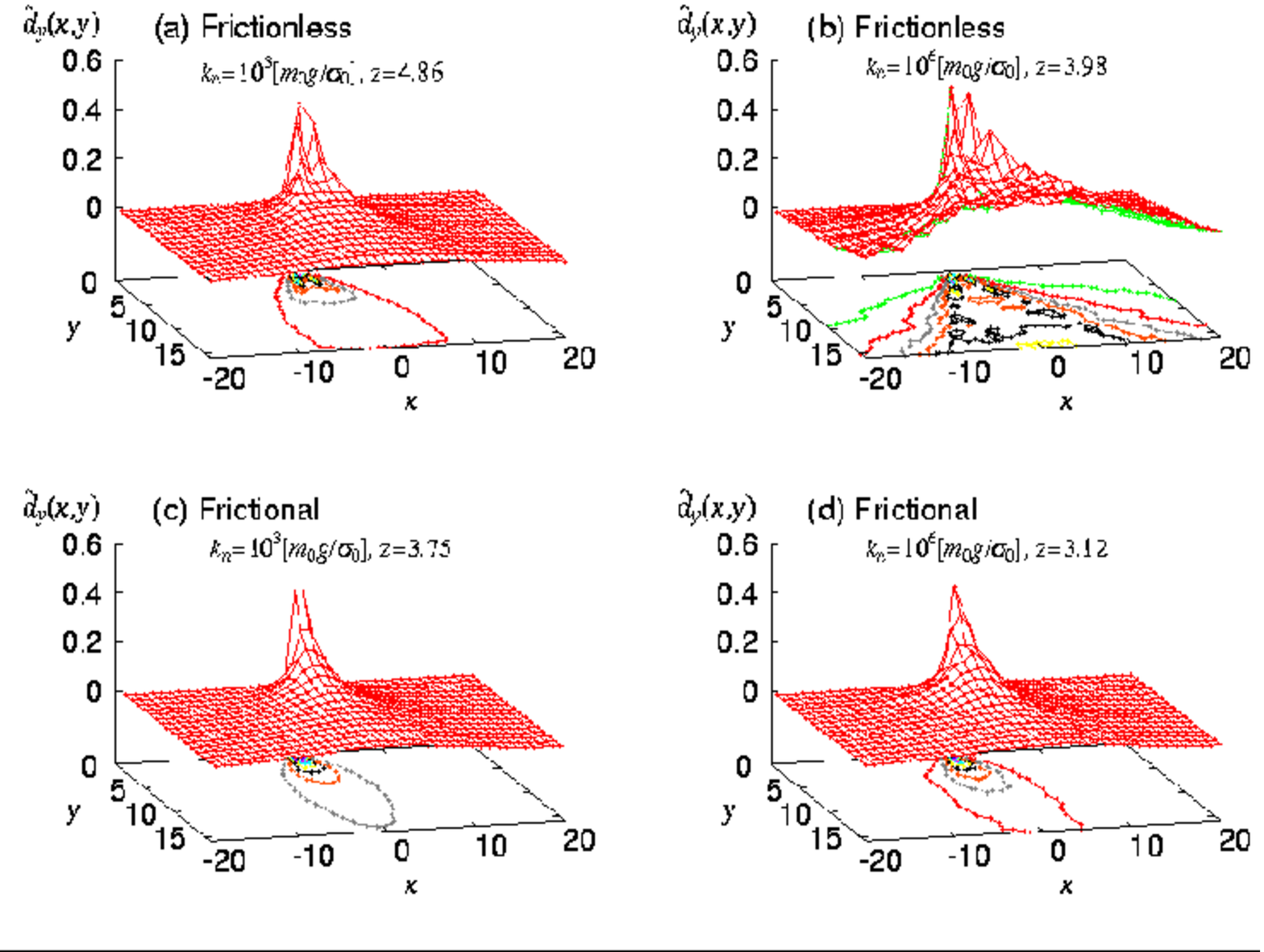}
\end{center}
\caption{(Color online)
The spatial distribution of the averaged absolute value of the
 longitudinal response in the frictionless piles (a, b), and
in the frictional piles (c, d) with $k_n=10^3$[$m_0 g/\sigma_0$] and
 $10^6$[$m_0 g/\sigma_0$].
The plots are averaged over a few hundreds realizations.
  } \label{f-7}
\end{figure}

The results are shown in Fig.\ref{f-7}, where $\hat d_y(\mib{r})$'s are
plotted against $x$ and $y$ with the contour lines;
The perturbation is applied at $(x,y)=(0,0)$.

We examine the two cases, namely, $k_n=10^3$ and $10^6$ [$m_0
g/\sigma_0$], for both the frictionless and frictional piles; the
coordination number $z=4.86$ and 3.98 for $k_n=10^3$ and $10^6$ [$m_0
g/\sigma_0$], respectively, for the frictionless pile, and $z=3.75$ and
3.12 for the frictional pile;  The piles with $k_n=10^6$ [$m_0
g/\sigma_0$] are closer to the isostatic than those with $k_n=10^3$
[$m_0g/\sigma_0$] for both the frictionless and frictional piles.

Let us examine the frictionless case first.  The response in the pile
with $k_n=10^3$[$m_0g/\sigma_0$], which is away from the isostatic,
decays quickly as it departs from the point of perturbation.  On the
other hand, the situation is quite different for the pile with
$k_n=10^6$ [$m_0 g/\sigma_0$], which is close to the isostatic: the
response does not decay along the $y$-axis with $x=0$, and if one looks
along the line parallel to the $x$-axis with constant $y$, one sees a
plateau region where the response is constant.  This plateau region,
which seems to extend to the infinity, is responsible to the diverging
behavior of the total response $D_y$.

On the other hand, in the case of frictional pile with $k_n=10^6$
[$m_0g/\sigma_0$], there is no tendency to develop the plateau region,
although the response is larger than that for the pile of $k_n=10^3$
[$m_0 g/\sigma_0$].

%------------------------------------
\subsection{Averaged Displacement-Displacement Response Function}

Finally, we present the displacement-displacement response function
$\mib{\Delta}(\mib{r})$ for this external perturbation:
\begin{equation}
\mib{\Delta}(\mib{r}) = <\sum_{i=1}^N \mib{d}_i \delta(\mib{r}-\mib{r}_i)>.
\end{equation}
The average is taken over a few hundreds realizations and the spatial
dependence is calculated on a grid with the mesh size $\sigma_0$.

The displacement-displacement response function has been shown to be
equal to the force-force response function for a frictionless isostatic
structure\cite{M01}, but it should be noted that this correspondence
does not hold in other cases.

\begin{figure}[tb]
\begin{center}
\includegraphics[width=17cm,clip]{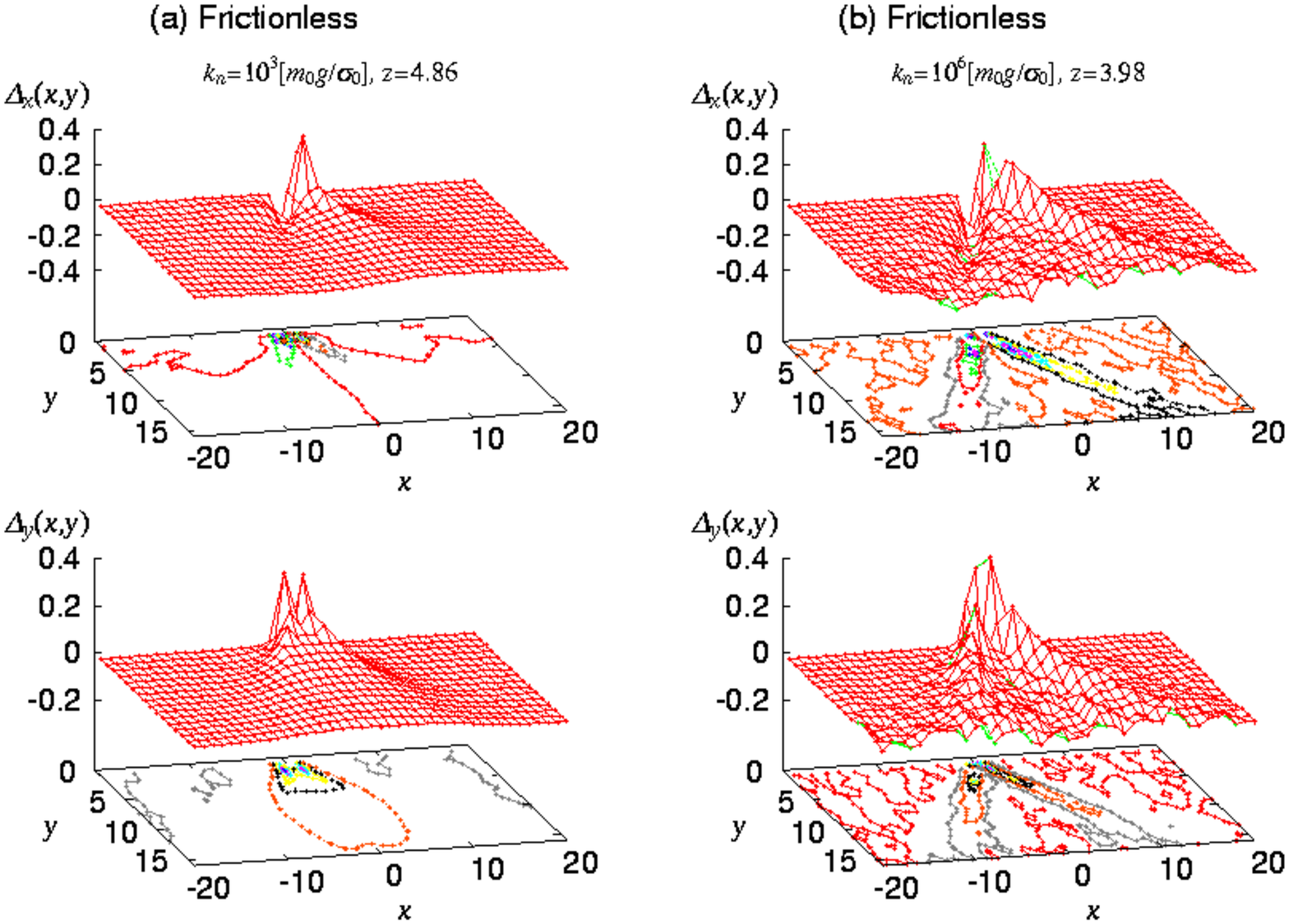}
\end{center}
\caption{(Color online)
The spatial distribution of the averaged displacement-displacement
response function
in the frictionless piles
 with $k_n=10^3$[$m_0g/\sigma_0$] (a) and
 $10^6$[$m_0g/\sigma_0$] (b).
The plots are averaged over a few hundreds realizations.
} \label{f-8}
\end{figure}

\begin{figure}[tb]
\begin{center}
\includegraphics[width=17cm,clip]{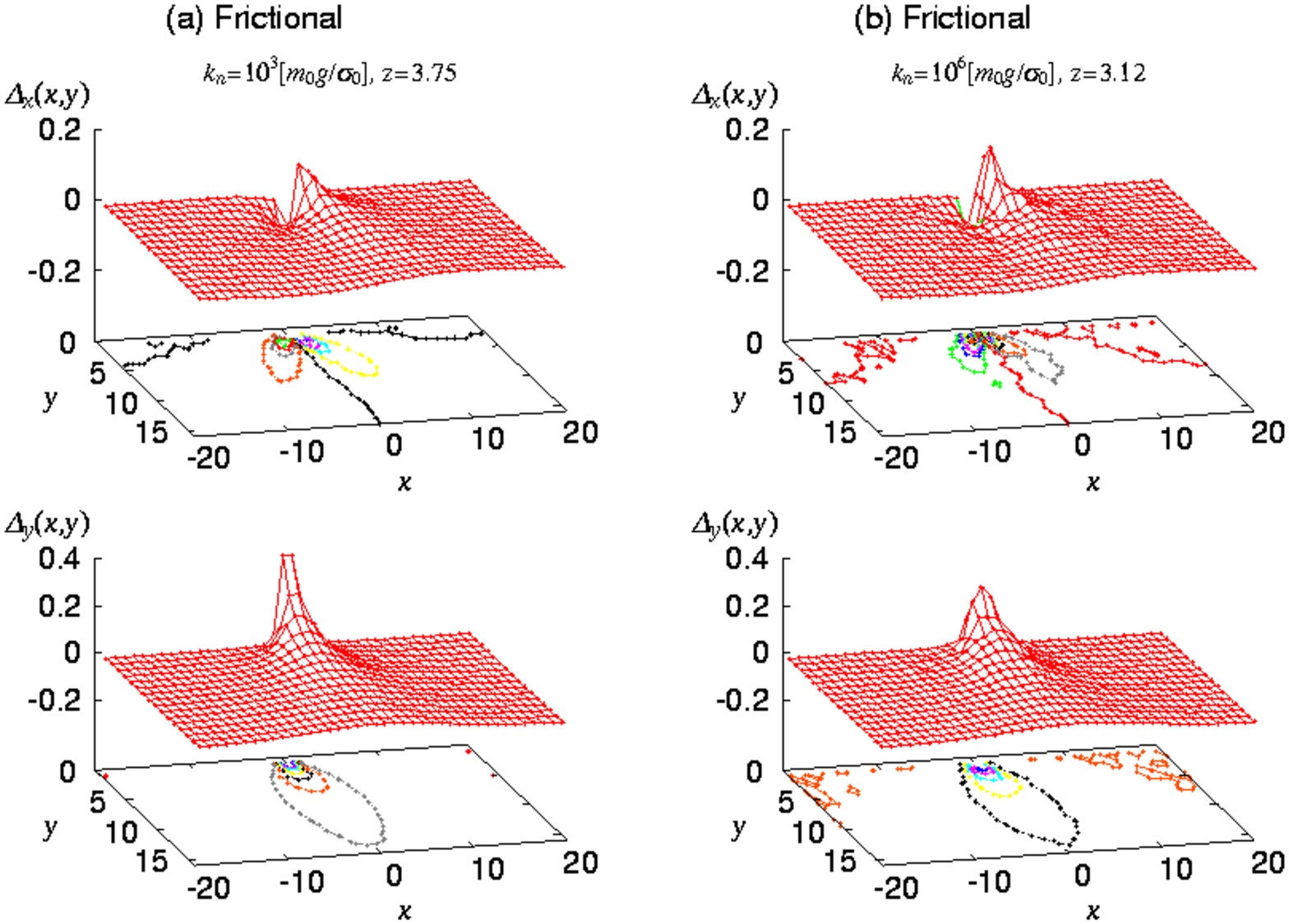}
\end{center}
\caption{(Color online)
The spatial distribution of the averaged displacement-displacement
response function
in the frictional piles
 with $k_n=10^3$[$m_0g/\sigma_0$] (a) and
 $10^6$[$m_0g/\sigma_0$] (b).
The plots are averaged over a few hundreds realizations.
} \label{f-9}
\end{figure}

The results are shown in Figs. \ref{f-8} and \ref{f-9} for the
frictionless and frictional piles, respectively, for $k_n=10^3$ and
$10^6$ [$m_0 g/\sigma_0$].

$\Delta_x(\mib{r})$ is positive for $x>0$ and negative for $x<0$ while
$\Delta_y(\mib{r})$ is always positive when it is averaged.

For both the frictionless and frictional piles, the region where
$\mib{\Delta}(\mib{r})\neq \mib{0}$ is larger for the pile whose coordination
number $z$ is smaller, namely, for the pile that is closer to the
isostatic.

It should be noted that
the way $\Delta_y(\mib{r})$ extends is very different from
that of $\hat d_y(\mib{r})$, especially in the frictionless pile close to
the isostatic.
The response function $\Delta_y(\mib{r})$ propagates in the
$y$-direction with the double peaked structure when one see it along the
line parallel to the $x$-axis with constant $y$, while
$\hat d_y(\mib{r})$ develops the plateau region.

This comparison shows that the response to the perturbation is not
actually small in the ``low response region'' of $\Delta_y(\mib{r})$
between the two peaks for the frictionless isostatic  pile, but the
large response varies from sample to sample and they are averaged out to
make $\Delta_y(\mib{r})$ small.

\begin{figure}[tb]
\begin{center}
\includegraphics[width=17cm]{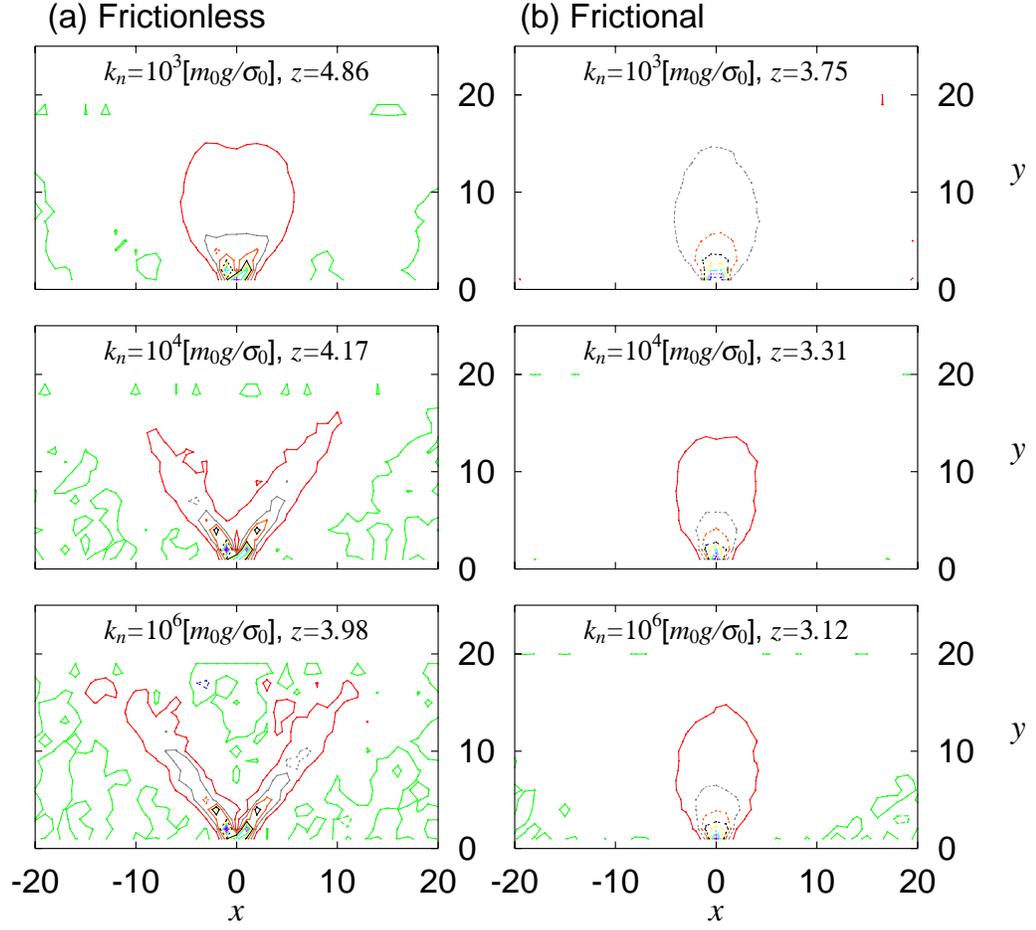}
\end{center}
\caption{(Color online)
The contour plots of the longitudinal component of
 displacement-displacement response function $\Delta_y$
in the frictionless piles (a) and in the frictional piles (b)
with the disk elasticity $k_n=10^3$, $10^4$, and $10^6$ [$m_0 g/\sigma_0$].
  } \label{f-10}
\end{figure}

The way how this double peaked structure in $\Delta_y(\mib{r})$ develops
as the frictionless pile approaches to isostatic can be seen in the
contour line plots in Fig.\ref{f-10} (a).

The response $\Delta(\mib{r})$ in the frictional pile shows smoother
structure than that in the frictionless pile and does not seems to
develop the double peaked structure as the pile becomes isostatic
(Fig.\ref{f-10} (b)).

%------------------ Summary and Discussions----------------
\section{Summary}

We have investigated the structure and the mechanical response of the
two-dimensional piles of disks with various disk elasticities; The piles
are formed through a several deposition procedures under the gravity.

As for the structure, we have shown the followings: The piles of
frictionless disks become isostatic when the disks are very hard and
they are not very sensitive to the preparation procedure, which is
consistent with the conjecture that the pile of rigid grains is
isostatic.  On the other hand, for the piles of frictional disks with
the infinite friction, the structure depends on the preparation process.
If the pile is formed from a triangular lattice with the inertia
following the Newtonian equation, the pile structure seems to be
distinctively different from the isostatic one even in the rigid limit,
as has been found in the previous work on the three-dimensional
system\cite{SEGHL02}.  We have found, however, that the pile of
frictional disks becomes very close to the isostatic one in the rigid
grain limit when we employ the deposition process where the infinite
friction is effective, namely, the viscous equation is used for the time
development for the disks with the infinite friction constant from the
random initial configurations.

The role of friction for the frictional isostaticity is demonstrated by
examining the distribution of the ratio of the normal force to the
tangential force; In the pile that is close to the isostaticity with
frictional disks, the distribution of $\zeta$ extends to very large
value as of order $10^3$.  This suggests that the frictional
isostaticity is realized only in the cases where the exceptionally large
friction coefficient is effective, and most of real stable piles with
modest friction should be hyperstatic with history dependent forces even
in the rigid limit.

We have also investigated the mechanical response of the piles of
frictionless and frictional disks with a special attention on the
isostaticity of the pile structure.  We have examined the disk
displacement caused by moving one of the disks at the bottom of the pile
by an infinitesimally small distance.  It has been shown that, for the
frictionless pile with the isostatic structure, the response does not
decay as it departs from the point of perturbation, and the total sum of
the disk displacement diverges as the pile becomes isostatic, while the
response decays quickly in the pile with structure far from
isostaticity.  It is found that the averaged longitudinal response
$\Delta_y$ becomes double peaked in the isostatic frictionless pile.

As for the frictional pile, the response function always decays in a
finite distance and the longitudinal response remains single peaked,
which behavior does not change drastically as the structure approaches
isostatic.

The double peaked structure in the displacement-displacement response
function for the frictionless isostatic pile may be compared with the
hyperbolic stress propagation in the granular system\cite{BCC95,WCC97}
due to the equivalence between the stress-stress response function and
the displacement-displacement response function in the frictionless
isostatic pile\cite{M01}.
The single peaked structure, on the other hand, may correspond to the
diffusive stress propagation, but this correspondence is indirect because
there is no equivalence between the displacement response
and the stress response in the frictional or non-isostatic pile.

\begin{figure}[b]
\begin{center}
%\includegraphics[width=5cm,angle=-90,clip]{fig-5a.eps} 
%\vskip 0.5cm
%\includegraphics[width=5cm,angle=-90,clip]{fig-5b.eps}
\includegraphics[width=8.6cm,angle=0,clip]{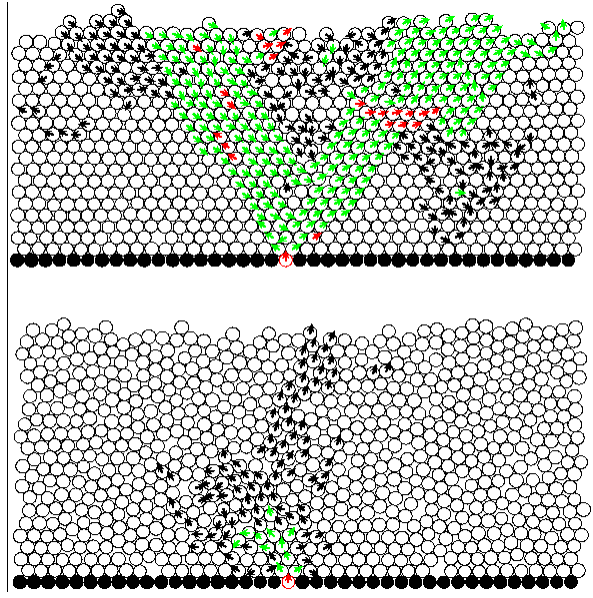}
\end{center}
\caption{(Color online) The mechanical response caused by the small
displacement of a disk at the bottom (denoted by an open (red) circle
with a dark grey (red) arrow) in the frictionless pile (upper) and in
the frictional pile (lower) with $k_n=10^6$[$m_0 g/\sigma_0$].  The dark
grey (red), light grey (green), and black allows denote the displacement
direction of the disks that move by the distance $\delta r_i \ge \delta
r_0$, $0.5\delta r_0 \le \delta r_i < \delta r_0$, and $0.1\delta r_0
\le \delta r_i < 0.5 \delta r_0$, respectively.  } \label{f-5}
\end{figure}

It is interesting to note that, in the isostatic piles, the spatial
distribution of averaged response of disk displacement $\Delta_y$ is
clearly different from that of absolute value of disk displacement;  the
former develops the double peaked structure while the latter shows the
plateau structure, namely, in the region between the two peaks where the
average displacement is small, the actual displacement response is not
small in each sample, but just they are random and averaged out.  This
means that the averaged displacement response in the
isostatic pile appears to propagate like a wave following a hyperbolic
equation, but the way that displacement response propagates in each
sample is quite complicated and does not look like a wave(Fig. \ref{f-5}).

%---------------------------------------
\begin{acknowledgments}
This work is partially supported by
Grant-in-Aid for Scientific Research (C) (No. 16540344) from 
JSPS, Japan.
\end{acknowledgments}

% the Ministry of Education, Culture, Sports, Science and Technology, Japan.

%-----------------------------------------------------------------------

%----------------------------------------------------------------

\newpage
\vspace*{5cm}
%----------------------------------------------------------------------
\begin{table}[h]
\begin{tabular}{|c|c|cc|cc|}
\hline
Eq. of motion & Initial config.  & 
\multicolumn{2}{c|}{Frictionless} &
\multicolumn{2}{c|}{Frictional} \\
& & $z_\infty$ & $\alpha$ & $z_\infty$ & $\alpha$ 
 \\
\hline
Newtonian & Triangular lattice & 3.97  & 0.68  & 3.15 & 0.49
\\
 & Random config. & 3.98 & 0.65 & 3.09 & 0.47
\\ \hline
Viscous & Triangular lattice & 3.97 & 0.63 & 3.06 & 0.60
\\
 & Random config. & 3.97 & 0.64 & 3.04 & 0.46
\\
%\cline{2-3}\\
\hline
\end{tabular}
\caption{
The limiting coordination numbers $z_\infty$ and the exponents $\alpha$
 for various preparation procedures.
}
\label{t-1}
\end{table}
%----------------------------------------------------------------------
.

%\newpage

%-----------------------------------------------
%-----------------------------------------------
\newpage
%-----------------------------------------------
%-----------------------------------------------
\newpage
%-----------------------------------------------
%-----------------------------------------------
%-----------------------------------------------

%-----------------------------------------------
%-----------------------------------------------

%-----------------------------------------------
%-----------------------------------------------

%-----------------------------------------------
%-----------------------------------------------

%-----------------------------------------------
%-----------------------------------------------

%-----------------------------------------------
%-----------------------------------------------

%-----------------------------------------------
%-----------------------------------------------

\end{document}